\documentclass{jfm}
\usepackage{amsmath}
\RequirePackage{lfmath} 
\usepackage[sort&compress,longnamesfirst,colon,round]{natbib}

\title[Net Upscale Energy Cascade]{A New Proof on Net Upscale Energy Cascade in 2D and QG Turbulence}
\author{Eleftherios Gkioulekas$^1$ and Ka Kit Tung$^2$}
\affiliation{$^1$Department of Mathematics, University  of Central Florida, Orlando, FL 32816-1364\\[\affilskip]
$^2$Department of Applied Mathematics, University of Washington, Seattle, WA 98195-2420}

\begin{document}
\maketitle

\begin{abstract}
A general proof that more energy flows upscale than downscale in two-dimensional (2D) turbulence and barotropic quasi-geostrophic (QG) turbulence is given.  A proof is also given that in Surface QG turbulence, the reverse is true.  Though some of these results are known in restricted cases, the proofs given here are pedagogically simpler, require fewer assumptions and apply to both forced and unforced cases.
\end{abstract}

\section{Introduction}

It is a well-known  result that energy is transferred by nonlinear wave-wave interaction predominantly upscale in two-dimensional homogeneous and isotropic (2D) turbulence, and in quasi-geostrophic (QG) turbulence (see \citet{book:Salmon:1998}).  What is less commonly known is the fact that, except for certain special  cases, a general unified proof spanning both the forced-dissipative and the decaying cases is not yet available. Furthermore, it is also not widely appreciated that these results cannot be readily extended to models of quasi-geostrophic turbulence. 

It was recognized by  \citet{article:Fjortoft:1953} and  \citet{article:Charney:1971}  that the direction of net energy transfer in 2D and QG turbulence may be different from that for 3D isotropic and homogeneous turbulence and that the cause for this different behavior should be attributed to the former's twin conservation of energy and enstrophy. However, as pointed out previously by  \citet{article:Warn:1975} and \citet{article:Welch:2001}, the analysis of triadic transfers by  \citet{article:Fjortoft:1953} and  \citet{article:Charney:1971}  was  flawed.  These proofs made use of the simultaneous conservation of energy and enstrophy in 2D and QG turbulence, and the fact that enstrophy spectrum $G(k)$ is related to the energy spectrum $E(k)$  by $G(k) = k^2 E(k)$.  They  claimed to have shown that if a unit of energy is moved downscale, many more units of it have to be moved upscale in order to preserve the twin energy and enstrophy conservation.  
However, the direction of energy flow in time cannot and should not be determined by conservational considerations alone. Either an essential use of the dissipation terms has to be made to set the direction of the time arrow, or a constraining assumption has to be introduced on the initial condition to employ a Boltzmann-type argument.


In his paper, \citet{article:Fjortoft:1953} gives two distinct proofs. The first proof by \citet{article:Fjortoft:1953} does show that the only admissible triad interactions  are those that spread energy from the middle wavenumber to the outer wavenumbers (and vice versa, the ones that bring in energy to the central wavenumber from the outer wavenumbers). These are the triad interactions defined by \citet{article:Waleffe:1992} as class ``R''. An alternative set of triad interactions are the ones where  energy is transfered from the smallest wavenumber to the two largest ones; these are the class ``F'' triad interactions, and they are dominant in three-dimensional turbulence. Fj{\o}rt{\o}ft's proof can be employed to rule these out in two-dimensional turbulence. \emph{However}, as was pointed out by  \citet{article:Warn:1975} there exist also class ``R'' triad interactions that transfer more energy downscale  than upscale. Thus, eliminating  the class ``F'' interactions is not sufficient to constrain the direction of the energy flux or the enstrophy flux. Despite this problem, Fj{\o}rt{\o}ft's proof has been popularized in textbooks \citep{book:Salmon:1998} and review articles \citep{article:Tabeling:2002} as a rigorous proof that constrains the direction of the fluxes in two-dimensional turbulence, thereby becoming a bit of a misunderstood ``folklore'' argument. 

The second result of \citet{article:Fjortoft:1953} is an upper bound on the total energy accumulated on wavenumbers larger than some given $k$. This result however applies only to initial value problems without forcing, where energy has to be bounded, unsurprisingly.  This inequality was later taken by \citet{article:Charney:1971} as a proof that energy cannot go downscale, since the  energy $E^{>k}(t)$ accumulated at wavenumbers larger than $k$ is bounded by 
\begin{equation}
E^{>k}(t) \leq \frac{1}{k^2} G (t) \leq \frac{1}{k^2}G(0)
\end{equation}
where $G(t)$ is the total enstrophy at time $t$. Thus, the energy spectrum $E(k,t)$ is bounded by $E(k,t) \leq c k^{-3}$ for some constant $c$. \citet{article:Welch:2001} pointed out that this behavior of the energy spectrum is merely a consequence of the requirement for convergence of the Fourier representation of the enstrophy spectrum $G(k)$, which implies that $G(k)$ must decay faster than $k^{-1}$ as $k \goto \infty$.  Therefore the energy spectrum $E(k)$ must decay faster than $k^{-3}$ as $k \goto \infty$.  It says nothing about the direction of energy cascade, thus it does not help Fj{\o}rt{\o}ft's ``proof'' in the first half of the paper. On the other hand, equation \eqref{eq:danilov} of our paper, which involves fluxes, is derived via the same mathematical trick that \citet{article:Fjortoft:1953} first used to derive his inequality, which involves spectra.

The deeper conceptual difficulty with Fj{\o}rt{\o}ft's  result was also recognized by  \citet{article:Kraichnan:1967:1} in section 3 of his paper.  As is well-known, the Euler equation is invariant with respect to time reversal, and as Kraichnan himself has observed, the direction of the energy and enstrophy flux can be reversed simply ``by reversing the velocity field everywhere in space''.  In other words, for every initial condition where the fluxes go one way, there is another initial condition where they go the opposite way. This conundrum is similar to the situation in statistical mechanics where experience suggests that entropy-increasing solutions are statistically more probable than the entropy-decreasing solutions even though the underlying dynamical system is symmetric under time reversal.  The arguments used to resolve this apparent paradox involve selecting a class of initial conditions which give entropy increasing solutions, and arguing in some way that the initial conditions have to be members of that class in order to be physically realistic. \citet{article:Kraichnan:1967:1} tries to define the direction of the time arrow by assuming  that the cascade energy spectrum has an ``urge'' to go towards the energy spectrum corresponding to absolute thermodynamic equilibrium, which implicitly assumes the validity of a Boltzmann-type thermodynamic argument.  Except for an interesting proof by \citet{article:Rhines:1975,article:Rhines:1979}, which will be discussed next, we are not aware of any convincing such argument that decides that the direction of the fluxes for the case of the Euler equation points toward thermodynamic equilibrium. On the other hand, because the Navier-Stokes equations are dissipative and more realistic than the Euler equation, a proof for the Navier-Stokes equation is more relevant.  As we shall see, in the forced-dissipative  case, the direction of the time arrow is decided by the dissipation terms of the Navier-Stokes equations without requiring any further assumptions.

The Rhines proof is applicable to  the case of unforced decaying turbulence \citep{article:Rhines:1975,article:Rhines:1979,book:Salmon:1998}.  In his argument, Rhines begins with the assumption that an initial peak of energy in the energy spectrum has the tendency to spread out.  This assumption constrains the set of  initial conditions and thus defines the direction of the time arrow.  Then, he shows that the energy weighted wavenumber, which represents the average location of the peak, will decrease in time therefore moving the peak to larger length scales.  From this, he argues that the energy therefore has a tendency to go upscale. Although Rhines originally intended the proof to apply to both the viscous case and the inviscid case, there was an error in \citet{article:Rhines:1979}, where the dissipation of energy was ignored while that of the enstrophy was kept.  Consequently the proof, in its published form, is valid only for the inviscid case. This problem was remedied by \citet{article:Scott:2001}, who gave a corrected proof for the case  of molecular diffusion and Ekman damping. In the present paper  we will extend the proof to the  case of hyperdiffusion. However, we will  also show that, curiously enough, the proof does not work for the case of hypodiffusion. 

It should be emphasized that Rhines' proof derives a statement involving the time derivative of the global integral of a quantity involving the energy spectrum. As such, it establishes a global tendency for the energy spectrum as a whole to shift toward smaller wavenumbers.  However, it would be incorrect  to draw conclusions on the behavior of the energy flux at local intervals of wavenumbers from a global result.  For example, one cannot conclude from this proof that the energy flux on the downscale side of the forcing range goes upscale.  Furthermore, because the scope of the proof is confined to the decaying problem, one can draw no conclusions on the direction of the energy flux from this proof for the forced-dissipative case.

A  nice proof was given by   \citet{article:Eyink:1996} for the forced-dissipative case in section 3.1.1 of his paper (see eqs. (3.2) to (3.8)). Similar proofs have also been given by \citet{submitted:KLB}, \citet{article:Danilov:2003} and \citet{article:Tung:2005:1}.  The advantage of this proof  is that it allows the dissipation terms to decide the direction of the time arrow.  Furthermore, it directly considers the behaviour of the energy flux instead of inferring it from the time-derivatives of the energy spectrum.  On the other hand, it cannot be easily extended to the case of decaying turbulence.  Furthermore, this proof \emph{also} requires an assumption: it requires  that inertial ranges exist and that there is a separation of scales between the upscale and downscale dissipation scale and the forcing scale. These assumptions are well supported by numerical simulations \citep{article:Vergassola:2000,article:Alvelius:2000,article:Kaneda:2001,article:Falkovich:2002}. However,  unlike the case of three-dimensional turbulence where the energy cascade is very robust, in two-dimensional turbulence there are situations where the inertial ranges do not exist \citep{article:Bowman:2003,article:Bowman:2004,article:Gurarie:2001,article:Gurarie:2001:1,article:Danilov:2003}.

In the remainder of his paper, \citet{article:Eyink:1996} shows that the underlying assumptions on the existence of inertial ranges can be rigorously reduced to the hypothesis that the total energy of the system remains finite in the limit $\nu\goto 0^{+}$ (see Hypothesis 1 in \citet{article:Eyink:1996}). As was pointed out to us by an anonymous referee, from a physical point of view, this hypothesis serves to rule out the possibility that the energy injected at the forcing range  will simply pile up in the spectral neighborhood of the injection and diverge to infinity in the limit $\nu\goto 0^{+}$. Thus, it is similar, in meaning, to the hypothesis of  \citet{article:Rhines:1975,article:Rhines:1979} that instead of piling up at forcing, the injected energy will want to spread out. From this hypothesis,  \citet{article:Eyink:1996} shows   that there exist regions of constant flux both upscale and downscale of the forcing scale and derives bounds on the location of the dissipation scales that prove a separation of scales between the dissipation scales and the forcing scale.  It should be noted  that the presence of a constant flux does not guarantee that a local cascade of energy and enstrophy is the dominant effect. A constant flux may also result from non-local transfer that takes energy and enstrophy directly from the forcing scale to the dissipation scales.

In the present paper we will use the energy flux approach and derive an inequality for the general case that shows that the weighted average of the energy flux is negative and the weighted average of the enstrophy flux is positive.  The averages involved are such that the inequalities can be satisfied only when most of the energy goes upscale and most of the enstrophy goes downscale.  For example, the energy flux inequality gives more weight to large wavenumbers than small wavenumbers.  Consequently, the upscale energy  flux at small wavenumbers must be significantly larger than the downscale flux at large wavenumbers to make the average come out negative.  A similar consideration applies to the enstrophy flux inequality.

What is remarkable is that in the forced-dissipative case these inequalities can be derived without any assumptions, except for requiring that the forcing spectrum is confined to a finite interval of wavenumbers, which can even be relaxed if necessary.  No assumptions on the existence of inertial ranges are necessary, which means that the inequalities are also valid in situations where the inertial ranges fail to exist.  For the case of decaying turbulence, it is necessary to make an assumption concerning the time derivative of the energy spectrum, but given that assumption the same inequalities continue to hold.  We believe that the reason why it is necessary  to make an assumption for the decaying case is  to weed out unusual initial conditions that might temporarily reverse the direction of fluxes.  In any event, the assumption involved is somewhat weaker than the assumption used by Rhines.

The paper is organized as follows. In section 2 we review the mathematical properties of the generalized one layer model.  The flux inequalities are proven for the forced-dissipative case in section 3. The implications for two-dimensional turbulence are discussed in section 4, and for models of quasi-geostrophic turbulence in section 5. A proof of the flux inequalities for the decaying case is given in section 6, and a review of the proof by Rhines in section 7.  The paper is concluded in section 8.  Appendix A reviews the H\" older inequalities, used in our discussion of the Rhines proof.



\section{Preliminaries}

 We shall first present the general case of the one-layer advection-diffusion model which encompasses 2D turbulence, CHM turbulence, and SQG turbulence, before considering the subcases separately. The governing equation of these systems has the distinguishing form of a conservation law for a vorticity-like quantity $\gz$:
\begin{equation}
\pderiv{\gz}{t} + J (\gy, \gz) = -[\nu (-\gD)^p + \nu_1 (-\gD)^{-h}] \gz + F,
\label{eq:governing}
\end{equation}
where  $\psi (x,y,t)$ is the streamfunction and $\gz$  is related to it through a linear operator $\cL$ by $\gz = -\cL \gy$. We assume that $\cL$ is a diagonal operator in Fourier space whose Fourier transform $L(k)$ satisfies $L(k)>0$ and $L'(k)>0$. The Jacobian term $ J (\gy, \gz)$ describes the advection of $\gz$ by $\gy$, and is defined as 
\begin{equation}
J (a,b) = \pderiv{a}{x} \pderiv{b}{y} - \pderiv{b}{x} \pderiv{a}{y}.
\end{equation}
  We have written the dissipation of $\gz$  in a more general form than normally used.  Our proof does not depend on the details of the operator $\cD = \nu (-\gD)^p + \nu_1 (-\gD)^{-h}$, only that it is a positive operator.  $F$  is the forcing function; $\nu$ is the hyperdiffusion coefficient; $\nu_1$ is the hypodiffusion coefficient. The physical case of molecular diffusion and Ekman damping corresponds to $p=1$ and $h=0$.

It can be shown that  if $a$ and $b$ satisfy a homogeneous (Dirichlet or Neumann) boundary condition, then $\snrm{J(a,b)} = 0$, where we use the notation  $\snrm{f}\equiv \iint(f(x,y))dxdy$. It follows from the product rule that
\begin{equation}
\snrm{J(ab,c)} = \snrm{aJ(b,c)} + \snrm{bJ(a,c)} = 0,
\end{equation}
from which we obtain the identity
\begin{equation}
\snrm{aJ(b,c)} = \snrm{bJ(c,a)} = \snrm{cJ(a,b)},
\end{equation}
which was also shown previously by \citet{submitted:KLB}. We assume that the operator $\cL$ is self-adjoint in the sense that it satisfies $\snrm{f (\cL g)} = \snrm{g (\cL f)}$ for any fields $f(x,y)$ and $g(x,y)$. This is true, if we assume that $\cL$ is diagonal in Fourier space. 

The conservation law $\pderivin{\gz}{t} + J(\gy,\gz) = 0$ conserves the ``enstrophy''-like quadratic $B = (1/2)\snrm{\gz^2}$ for any arbitrary linear operator $\cL$, because
\begin{equation}
\snrm{\dot B} = \snrm{\gz\dot\gz} = \snrm{-\gz J(\gy,\gz)} = \snrm{-\gy J(\gz, \gz)} = 0.
\end{equation}
For self-adjoint operators $\cL$, the ``energy''-like quadratic $A = (1/2)\nrm{-\gy\gz}$ is also conserved. To show that, note that
\begin{align}
\snrm{\dot A} &= (1/2)\snrm{-\gy\dot\gz - \gz\dot\gy} = (1/2)[\snrm{\gy J(\gy,\gz)} +  \snrm{\gz\cL^{-1}J(\gy,\gz)}] \\
&=  (1/2)[\snrm{\gy J(\gy,\gz)} + \snrm{\cL^{-1}\gz)J(\gy,\gz)}] \\
&=  \snrm{\gy J(\gy,\gz)}  = \snrm{\gz J(\gy,\gy)} = 0.
\end{align}
Let $A(k)$ and $B(k)$ be the spectral density of $A$ and $B$, respectively such that $A=\int_0^{+\infty} A(k) \; dk$ and $B=\int_0^{+\infty} B(k) \; dk$, and $k$ is the isotropic 2D wavenumber. The spectral equations are obtained by differentiating $A(k)$ and $B(k)$ with respect to $t$, and employing the Fourier transform of the governing equation \eqref{eq:governing}: 
\begin{align}
\frac{\partial A (k)}{\partial t}  + \frac{\partial \Pi_A (k)}{\partial k}  &=
-D_A (k) + F_A (k)\label{eq:consA}\\
\frac{\partial B (k)}{\partial t}  + \frac{\partial \Pi_B (k)}{\partial k}  &=
-D_B (k) + F_B (k)\label{eq:consB}.
\end{align}
 It is understood that ensemble averages have been taken in the above quantities. Here $\Pi_A (k)$ is the spectral density of $A$ transfered from $(0,k)$ to $(k,+\infty)$ per unit time by the nonlinear term in \eqref{eq:governing}, $D_A (k)$ the  dissipation of $A$, and $F_A (k)$ the  forcing spectrum of $A$, and likewise for the $B$ equation. The conservation laws imply for the viscous case that $\Pi_A (0) = \lim_{k\goto\infty} \Pi_A (k) = 0$ and  $\Pi_B (0) = \lim_{k\goto\infty} \Pi_B (k) = 0$. For the inviscid case, this condition can be violated, in principle, by anomalous dissipation for solutions that have singularities. The spectra of $A$ and $B$ are related as $B(k)=L(k)A(k)$, and likewise it is easy to show, from the diagonal structure of the $\cL$ operator in Fourier space, that $D_B(k)=L(k)D_A (k)$ and  $F_B (k)=L(k)F_A(k)$. Combining these equations with \eqref{eq:consB} and \eqref{eq:consA} we obtain the so-called Leith constraint \citep{article:Leith:1968}:
\begin{equation}
\pderiv{\Pi_B (k)}{k} = L(k) \pderiv{\Pi_A (k)}{k},
\end{equation}
which shows that if $\Pi_B (k)$ is strictly constant, then $\Pi_A (k)$  is also strictly constant and vice versa.

\section{Flux inequalities for the forced-dissipative case}

Assume that the forcing spectrum $F_A (k)$ is confined to a narrow interval of wavenumbers $[k_1, k_2]$. Then, we have
\begin{equation}
F_A (k) = 0 \text{ and } F_B (k) = 0, \forall k\in (0, k_1)\cup (k_2, +\infty), \label{eq:forceassumption}
\end{equation}
and we can show, without making any ad hoc assumptions, that under stationarity, the fluxes $\Pi_A (k)$ and $\Pi_B (k)$ will satisfy the inequalities
\begin{align}
\int_0^k L'(q)  \Pi_A (q) \; dq &<0,  \;\forall k > k_2 \label{eq:ineqA}\\
\int_k^{+\infty} \frac{L'(q)}{[L (q)]^2} \Pi_B (q) &> 0, \;\forall k < k_1 \label{eq:ineqB}.
\end{align}

The $\Pi_A (k)$ inequality is shown as follows:
Integrating \eqref{eq:consA} and \eqref{eq:consB} over the $(k,+\infty)$ interval and employing the stationarity conditions $\pderivin{A(k)}{t} = 0$ and  $\pderivin{B(k)}{t} = 0$  gives:
\begin{align}
\Pi_A (k) &= \int_k^{+\infty} [D_A (q) - F_A (q)] \; dq \\
\Pi_B (k) &= \int_k^{+\infty} [D_B (q) - F_B (q)] \; dq =  \int_k^{+\infty} L(q) [D_A (q) - F_A (q)] \; dq.
\end{align}
Using integration by parts, and the Leith constraint, we have the relation
\begin{align}
 \Pi_B (k) &=  \int_0^k \pderiv{\Pi_B (q)}{q} \; dq = \int_0^k  L(q) \pderiv{\Pi_A (q)}{q}  \; dq \\
&= L(k)\Pi_A (k) - \int_0^k L'(q)  \Pi_A (q) \; dq,
\end{align}
 from which we obtain the inequality itself
\begin{align}
\int_0^k L'(q)  \Pi_A (q) \; dq &= L(k)\Pi_A (k) - \Pi_B (k) \label{eq:partsA}  \\
&= \int_k^{+\infty} [L(k) - L(q)] [D_A (q) - F_A (q)]\; dq \\
&< 0, \;\forall k\in (k_2,+\infty).
\label{eq:dissintA}
\end{align}
Here we use $L(q)-L(k)>0, \;\forall q\in (k,+\infty)$, and $D_A (q) - F_A (q)\geq 0$ which follows from $D_A(q)\geq 0$ and $F_A (q) = 0,  \forall q>k>k_2$.

The counterpart inequality for the flux $\Pi_B (k)$ can be  derived similarly. We begin by integrating \eqref{eq:consA} and \eqref{eq:consB}, but this time over the $(0,k)$ interval:
\begin{align}
\Pi_A (k) &= -\int_0^k  [D_A (q) - F_A (q)] \; dq \\
\Pi_B (k) &= -\int_ 0^k [D_B (q) - F_B (q)] \; dq =  -\int_0^k L(q) [D_A (q) - F_A (q)] \; dq.
\end{align}
Similarly, to avoid the singularity at $q=0$, we do the integration by parts over the  $(k,+\infty)$ interval:
\begin{align}
\Pi_A (k) &= - \int_k^{+\infty} \pderiv{\Pi_A (q)}{q} \; dq = -\int_k^{+\infty}  \frac{1}{L(q)} \pderiv{\Pi_B (q)}{q} \; dq\\
& = \frac{\Pi_B (k)}{L(k)} - \int_k^{+\infty} \frac{L'(q)}{[L (q)]^2} \Pi_B (q) \; dq,
\end{align}
and consequently, we obtain
\begin{align}
\int_k^{+\infty} \frac{L'(q)}{[L (q)]^2} \Pi_B (q) &= -\frac{L(k)\Pi_A (k) - \Pi_B (k)}{L(k)} \label{eq:partsB} \\
&= \frac{1}{L(k)}  \int_0^k [L(k) - L(q)] [D_A (q) - F_A (q)]\; dq \\
&> 0, \;\forall k\in (0,k_1). \label{eq:dissintB}
\end{align}
Here, the inequality changes direction, because $L(k)-L(q)>0,\;\forall q<k$.

Note that both proofs are based on the inequality
\begin{equation}
L(k)\Pi_A (k) - \Pi_B (k) < 0, \;\forall k\in (0, k_1)\cup (k_2, +\infty),
\label{eq:danilov}
\end{equation}
which holds both upscale and downscale of the forcing range in the forced-dissipative case discussed here. We called this inequality, in a previous paper \citep{article:Tung:2005:1}, the ``Danilov inequality'' because it was communicated to us by Danilov. It is worth noting that this inequality is the flux analog of a similar but distinct  inequality derived by \citet{article:Fjortoft:1953} in terms of the energy spectrum and the enstrophy spectrum. Finally, similar flux inequalities were known to \citet{article:Eyink:1996}. Eq. \eqref{eq:danilov} is a  sharper and more general variation of these previous results.

Also note that, for $k<k_1$, since $D_A (k) > 0$ for all $k$, it follows immediately from the steady state version of \eqref{eq:consA} and the assumption \eqref{eq:forceassumption} that 
\begin{equation}
\Pi_A (k) = -\int_0^k D_A (q) \; dq <0, \forall k\in (0, k_1).
\end{equation}
In general, one can easily show, for the forced-dissipative case where the forcing spectrum obeys \eqref{eq:forceassumption}, under statistical equilibrium, that
\begin{align}
\Pi_A (k) > 0 &\text{ and } \Pi_B (k) > 0, \;\forall k\in (k_2, +\infty) \\
\Pi_A (k) < 0 &\text{ and } \Pi_B (k) < 0, \;\forall k\in (0, k_1).
\end{align}
It follows that, contrary to some popular misconceptions, both fluxes go downscale on the downscale side of injection, and upscale  on the upscale side of injection.

\section{Implications for two-dimensional turbulence}

For the case of 2D turbulence, $A(k)$ is the energy spectrum $E(k)$, $B(k)$ is the enstrophy spectrum $G(k)$ and $L(k)=k^2$. The inequality \eqref{eq:ineqA} simplifies to:
\begin{equation}
\int_0^k 2q \Pi_E (q) \;dq < 0,\;\forall k\in (k_2, +\infty).
\label{eq:en2d}
\end{equation}
This integral constraint  implies that energy fluxes upscale in the net.   The constraint \eqref{eq:en2d} also holds trivially for $k<k_1$,  since $\Pi_A (k) < 0$ for all $k<k_1$.  For $k>k_2$, the integration range also includes the energy injection interval $[k_1, k_2]$ and both the upscale cascade range and the downscale cascade range.  The inequality \eqref{eq:en2d} implies that the negative flux in the $(0,k_1)$ interval is more intense than the positive flux in the $(k_2, +\infty)$ because the weighted average of $\Pi_E (k)$ gives more weight to the large wavenumbers.

Similarly, \eqref{eq:ineqB} reduces to
\begin{equation}
\int_k^{+\infty} 2q^{-3} \Pi_G (q) \; dq > 0,\;\forall k\in (0,k_1),
\label{eq:enst2d}
\end{equation}
which  is a statement that enstrophy fluxes downscale in the net.

The inequalities \eqref{eq:en2d} and \eqref{eq:enst2d} are the two main results in 2D turbulence we were looking for, and they constitute proofs that in forced-dissipative 2D turbulence under statistical equilibrium energy predominantly is transferred upscale while enstrophy downscale.

To understand the implications of these inequalities on two-dimensional turbulence we have to distinguish between the following cases and consider them separately:

(a) \emph{No infrared sink of energy, finite box:}
This is the case considered by \citet{article:Shepherd:2002}. The coefficient of hypoviscosity, which provides the sink at the large scales, is zero. i.e.  $\nu_1=0$.  The only dissipation mechanism is a very small molecular viscosity $\nu$, with $p=1$.  Our result of net energy cascade \eqref{eq:en2d} still holds.  However, without a sink of energy at large scales, the energy which is fluxed upscale piles up until it is dissipated by the small viscosity at the forcing scale \citep{article:Bowman:2003,article:Bowman:2004}.  No inertial range exists where the fluxes of energy and enstrophy are constant.  Nevertheless,   \eqref{eq:en2d} implies that there is more energy flux dissipated on the upscale side of the forcing range than on the downscale side of the forcing range, and likewise \eqref{eq:enst2d} implies that there is more enstrophy dissipated on downscale side of the forcing range than on the upscale side of the forcing range . 

(b) \emph{No infrared sink of energy, infinite box:}
Same as in case (a) except that the domain is infinite.  This is the classical case of 2D turbulence considered by \citet{article:Kraichnan:1967:1}, \citet{article:Leith:1968}, and \citet{article:Batchelor:1969}.  Although there is no infrared sink of energy, the energy cascaded upscale can keep on cascading to ever larger  scales.  There is no pile up of energy, but there is always a spectral region at larger and larger scales where steady state cannot be achieved.  Let this region be denoted by $0<k<k_0 (t)$. Quasi-steady state can be achieved for $k>k_0$.  In this latter spectral region, our inequalities \eqref{eq:en2d} and \eqref{eq:enst2d} do hold.  Since energy transferred upscale through $k_0$ is ``lost'' to the region downscale from $k_0$, the infinite domain acts in effect like a perfect infrared sink.  Furthermore, in the original formulation of the  KLB theory, the molecular viscosity coefficient  $\nu$ is taken to $\nu\goto 0^{+}$, with the result that the energy dissipated at the ultraviolet end of the spectrum vanishes in the limit.  In this configuration, all injected energy is transferred upscale and all injected enstrophy is transferred downscale.
These results for the KLB theory have been summarized by \citet{article:Tung:2005,article:Tung:2005:1}. 

(c) \emph{Finite infrared and ultraviolet sinks of energy:}
When there is a finite infrared sink of energy upscale of injection and a finite ultraviolet sink of energy downscale of injection, there is in general both an upscale and a downscale flux of energy.  This situation has been considered in  \citet{article:Eyink:1996} and \citet{article:Tung:2005,article:Tung:2005:1}. The upscale flux should be larger than the downscale flux, according to \eqref{eq:en2d}.  It should be noted that, because of the  inequality \eqref{eq:danilov}, the contribution of downscale energy flux to the energy spectrum in the inertial range on the downscale side of injection is always subleading and hidden.  This is not true in some baroclinic cases of QG turbulence \citep{submitted:Gkioulekas:3,lfthesis:2006}.

\section{Implications for models of QG turbulence}

As derived by \citet{article:Charney:1971}, QG turbulence conserves two quantities, total energy, which consists of horizontal components of kinetic energy plus available potential energy, and potential enstrophy. We now discuss briefly the implications of the flux inequalities on one-layer and two-layer simplifications of the quasi-geostrophic turbulence model.

(a) \emph{CHM turbulence:} This model is a two-dimensional version of the quasi-geostrophic model, and represents physically two-dimensional turbulence on a rotating frame of reference.  The governing equation is \eqref{eq:governing}  with $L(k)=k^2 + \gl^2$, where $\gl$ is the deformation wavenumber.  The total energy $E$ and total potential enstrophy $G$ are given by $E = (1/2)\nrm{|\del\psi |^2 + \gl^2 |\psi |^2}$ and $G = (1/2)|\gz |^2$. The flux inequalities are 
\begin{align}
\int_0^k 2q \Pi_E (q) \;dq &< 0,\;
\forall k\in (k_2, +\infty) \\
\int_k^{+\infty} \frac{2q}{(q^2 + \gl^2)^2} \Pi_G (q) \; dq &>0, \;\forall k\in (0, k_1),
\end{align}
 and they still imply that the total energy is mainly transferred upscale whereas the potential enstrophy is mainly transfer downscale.

(b) \emph{SQG turbulence:} This model can be derived from the quasi-geostrophic model by assuming that the potential vorticity is zero over the entire three-dimensional domain.  Then, it can be shown that the behaviour of the entire system is coupled to its behaviour in the boundary condition at the layer $z=0$ \citep{article:Orlando:2003:1}. At $z=0$, the potential temperature $\Theta$ is governed by  \eqref{eq:governing}  with $L(k)=k$, where $\Theta=\gz$. The conserved quadratic $B$ represents the total energy $E_{2D}$ of the system at the layer $z=0$, whereas the quadratic $A$ is the total energy $E_{3D}$ integrated over the whole domain $z\in (0,+\infty)$ \citep{submitted:Gkioulekas:3,lfthesis:2006}. In this system, there is no enstrophy, since the potential vorticity has been taken   equal to zero, and consequently there is no enstrophy cascade. The flux inequalities are 
\begin{align}
\int_k^{+\infty} & q^{-2} \Pi_{E_{2D}} (q) \; dq > 0, \;\forall k\in (0,k_1) \\
\int_0^k & \Pi_{E_{3D}} (q) \; dq < 0, \;\forall k\in (k_2, +\infty),  
\end{align}
and they imply that downscale from injection the dominant process is a downscale energy cascade at the layer $z=0$. Upscale from injection the energy spectrum is dominated by an inverse energy cascade of the total energy over the entire domain.  It should be noted that just as in two-dimensional turbulence, a dissipation sink is probably needed both upscale and downscale of injection to allow either cascade to form successfully.

(c) \emph{2-layer model of QG turbulence:} This model consists of two symmetrically coupled layers of two-dimensional turbulence where the deformation  wavenumber $\gl$ is the coupling constant \citep{article:Salmon:1978,article:Salmon:1980,book:Salmon:1998}. For the general baroclinic case, specifically with Ekman damping only in the lower layer, Danilov's inequality \eqref{eq:danilov} does not necessarily hold for 2-layer models (see \citet{submitted:Gkioulekas:3,lfthesis:2006}).  We therefore do not have a conclusive proof for the case of 2-layer models.  However, numerical results (see e.g. \citet{article:Orlando:2003}) show that most of the energy will still go upscale in this system, although some small fraction goes downscale.  In particular, the upscale energy cascade in the inertial range upscale of injection is much larger than the downscale flux of energy in the inertial range downscale of injection.

\section{Flux inequalities for the time-dependent case}

We now generalize the proof to time-dependent cases.  Since \eqref{eq:partsA} and \eqref{eq:partsB} are mathematical identities, they hold whether or not the quantities involved are time-dependent.
\begin{align}
\int_0^k L'(q)  \Pi_A (q) \; dq &=  L(k)\Pi_A (k) - \Pi_B (k),\;
\forall k\in (k_2, +\infty)\\
\int_k^{+\infty} \frac{L'(q)}{[L (q)]^2} \Pi_B (q) &= -\frac{L(k)\Pi_A (k) - \Pi_B (k)}{L(k)}, \;\forall k\in (0, k_1).
\end{align}
Equations \eqref{eq:dissintA} and \eqref{eq:dissintB}, however, should be modified to:
\begin{align}
L(k)\Pi_A (k) - \Pi_B (k) &= \int_k^{+\infty} [L(k) - L(q)] [D_A (q)-F_A (q) +\pderiv{A(q)}{t}]\; dq , \;\forall k\in (0, k_1)\\
L(k)\Pi_A (k) - \Pi_B (k) &= -\int_0^k  [L(k) - L(q)] [D_A (q)-F_A (q) +\pderiv{A(q)}{t}]\; dq,\;
\forall k\in (k_2, +\infty).
\end{align}
Choosing $k$ to be outside the forcing range $[k_1, k_2]$, and combining the previous four equations we obtain:
\begin{align}
\int_0^k L'(q)  \Pi_A (q) \; dq &= \int_k^{+\infty} [L(k) - L(q)] [D_A (q) +\pderiv{A(q)}{t}]\; dq ,\;
\forall k\in (k_2, +\infty) \label{eq:dineqA}\\
\int_k^{+\infty} \frac{L'(q)}{[L (q)]^2} \Pi_B (q) &= \frac{1}{L(k)}  \int_0^k [L(k) - L(q)] [D_A (q)  +\pderiv{A(q)}{t}]\; dq , \;\forall k\in (0, k_1). \label{eq:dineqB}
\end{align}
The equations \eqref{eq:dineqA} and \eqref{eq:dineqB} together are a general and remarkable result, because they relate the weighted mean of flux of $A$ in $(0,k)$ to what happens outside this range, and the weighted mean of flux of $B$ in  $(k,+\infty)$ to what happens outside $(k,+\infty)$.  

(a) \emph{Initial stage:}                       
 During the initial development, nonlinear interactions transfer energy from one wavenumber to another.  If the initial condition $A_0 (k)$  for $A(k)$ is of compact support (which is almost always the case in reality) then we can expect that during the initial stages of decay where $A$ is still in the process of spreading  there will be a small wavenumber $\gee_1 >0$ and a large wavenumber $\gee_2 > 0$ such that
\begin{equation}
A_0 (k)=0 \text{ and } \pderiv{A(k)}{t} \geq 0 ,\;\forall k\in (0, \gee_1)\cup (\gee_2, +\infty).
\end{equation}
Combining this condition with \eqref{eq:dineqA} and \eqref{eq:dineqB}, it follows that:
\begin{align}
\int_0^k L'(q)  \Pi_A (q) \; dq &\leq \int_k^{+\infty} [L(k) - L(q)] \pderiv{A(q)}{t}\; dq \leq 0 ,\;
\forall k\in (\gee_2, +\infty) \label{eq:ddineqA} \\
\int_k^{+\infty} \frac{L'(q)}{[L (q)]^2} \Pi_B (q) &\geq \frac{1}{L(k)}  \int_0^k [L(k) - L(q)] \pderiv{A(q)}{t}\geq 0\; dq   , \;\forall k\in (0, \gee_1).\label{eq:ddineqB}
\end{align}
Note that each of the two previous inequalities uses only part of the assumption, i.e.
\begin{align}
\pderiv{A(k)}{t} > 0 ,\;\forall k\in (\gee_2, +\infty)& \implies \int_0^k L'(q)  \Pi_A (q) \; dq < 0,\;\forall k\in (\gee_2, +\infty) \\
\pderiv{A(k)}{t} > 0 ,\;\forall k\in(0, \gee_1)&\implies\int_k^{+\infty} \frac{L'(q)}{[L (q)]^2} \Pi_B (q)>0  ,\;\forall k\in(0, \gee_1).
\end{align}
Furthermore, for $t=0$, the assumption 
\begin{equation}
A_0 (k)=0, \;\forall k\in (0, \gee_1)\cup (\gee_2, +\infty),
\end{equation}
 implies that 
\begin{equation}
\left. \pderiv{A(k)}{t}\right|_{t=0} \geq 0, \;\forall k\in (0, \gee_1)\cup (\gee_2, +\infty),
\end{equation}
from the positivity of $A(k)$. However for $t>0$ the latter is an additional hypothesis.

(b) \emph{Intermediate stage:}
In the intermediate stage, nonlinear spreading and dissipation are both active at the small scales.  Nonlinear transfer still supplies some $A$ to small and large scales by spreading.  Therefore  
\begin{equation}
D_A (q) +\pderiv{A(q)}{t}\geq 0,\;\forall k\in (0, \gee_1)\cup (\gee_2, +\infty),
\end{equation}
 and so from \eqref{eq:dineqA} and\eqref{eq:dineqB}) we again obtain 
\begin{align}
D_A (q) +\pderiv{A(k)}{t}\geq  0 ,\;\forall k\in (\gee_2, +\infty)& \implies \int_0^k L'(q)  \Pi_A (q) \; dq \leq 0,\;\forall k\in (\gee_2, +\infty) \\
D_A (q) +\pderiv{A(k)}{t} \geq 0 ,\;\forall k\in(0, \gee_1)&\implies\int_k^{+\infty} \frac{L'(q)}{[L (q)]^2} \Pi_B (q)\geq 0  ,\;\forall k\in(0, \gee_1).
\end{align}

(c)  \emph{Final decaying stage:}
In the final stages of unforced turbulence, $A$ decays due to dissipation.  The decay rate of $A(k)$ is the same as the dissipation rate.  Therefore,
\begin{equation}
D_A (q) +\pderiv{A(q)}{t}= 0,\;\forall k\in (0, \gee_1)\cup (\gee_2, +\infty),
\end{equation}
and consequently,
\begin{align}
\int_0^k L'(q)  \Pi_A (q) \; dq = 0,\;\forall k\in (\gee_2, +\infty)\\
\int_k^{+\infty} \frac{L'(q)}{[L (q)]^2} \Pi_B (q)=0,\;\forall k\in(0, \gee_1).
\end{align}
 We do not have upscale cascade.  During this final stage, nonlinear spreading has already occurred, and dissipation of energy dominates.   Nevertheless, this still implies that $A$ is transferred in the net upscale and $B$ in the net downscale.

The implication of these results is that  net energy flux is directed in the net upscale  for the time-dependent case of 2D and barotropic QG turbulence in the absence of forcing if the initial condition is of compact support and if it is assumed that it subsequently spreads into small scales. For SQG turbulence the result is reversed, in the sense that energy in the $z=0$ layer is transferred downscale in the net.

\section{Remarks on Rhines proof}
\label{sec:rhines}

Rhines starts with the assumption:
\begin{equation}
\frac{d}{dt}\int_0^{+\infty} (k-K)^2 E(k) \; dk > 0,
\label{eq:rhinesass}
\end{equation}
where $K = E_1/ E_0$ is the first moment of $E(k)$ and $E_a$ is defined as
\begin{equation}
E_a = \int_0^{+\infty} k^a E(k) \; dk.
\end{equation}
 Here \eqref{eq:rhinesass} is a ``postulate that the peak will spread in time'' from its current centre of ``mass'' $K$ \citet{article:Rhines:1975}, not necessarily in particular realizations, but in a probabilistic sense where an ensemble average over all initial conditions, constrained by the initial energy spectrum, has been taken.  Rhines then shows that 
\begin{equation}\dderiv{K^2}{t} <0,\end{equation}
 which means that the average location of the peak tends to move toward smaller wavenumbers, and concludes from this that the energy has a tendency to be transfered  upscale.

In \citet{article:Rhines:1975}, the details of the proof  are not given.  In \citet{article:Rhines:1979}, the following more detailed argument is given which is correct for the inviscid case  $\nu=0$ and $\nu_1=0$: Expanding 
\begin{equation}
\int_0^{+\infty} (k-K)^2 E(k) \; dk = E_2 - 2K E_1 + K^2 E_0 = E_2 - K^2 E_0,
\end{equation}
and solving for $K^2$, we obtain
\begin{equation}
E_0 K^2 =E_2 -  \int_0^{+\infty} (k-K)^2 E(k) \; dk.
\end{equation}
Differentiating with respect to $t$, and writing $E_a' = dE_a/dt$, we have,
\begin{equation}
E_0' K^2 + E_0 \frac{dK^2}{dt} = E_2' - \frac{d}{dt}\int_0^{+\infty} (k-K)^2 E(k) \; dk < E_2',
\label{eq:rhines}
\end{equation}
which gives,
\begin{equation}
\frac{dK^2}{dt} < \frac{E_2' -E_0' K^2}{E_0}.
\label{eq:rhines2}
\end{equation}
If we assume $E'_0 = 0$ and $E'_2=0$, which can be deduced from conservation of energy and enstrophy for the case where there are no viscosities, then it follows that \begin{equation}
\dderiv{K^2}{t} <0. 
\end{equation}

However, Rhines' argument was supposed to work for the viscous case as well (see page 405, last equation, of \citet{article:Rhines:1979}) where $E'_0 < 0$ and $E'_2< 0$.  It appears that the term $E'_0 K^2$ in  \eqref{eq:rhines} was ignored in that derivation.  If the term is included, then the right hand side of \eqref{eq:rhines2} has two terms of opposite sign, and it is not immediately clear which term dominates.  Nevertheless, \citet{article:Scott:2001} showed, using the Holder inequality, that the proof can still be completed, for the case of Ekman damping and molecular diffusion ($h=0$ and $p=1$). 

As it stands, this proof is interesting, but it cannot be extended to the forced-dissipative case because it relies on describing the behavior of time-derivatives of the energy spectrum rather than fluxes. Furthermore, it relies on the  assumption \eqref{eq:rhinesass}, without proof. The difference between the assumption \eqref{eq:rhinesass} and the assumption used in our proof is that, \eqref{eq:rhinesass} is a global condition stated over the entire range of wavenumbers, whereas the assumption needed for our proof in the previous sections is a local condition over the intervals $(0, \gee_1)\cup (\gee_2, +\infty)$. We suspect that the need to make some assumption for proofs covering the decaying case is unavoidable because it is  necessary to weed out unusual initial conditions.

 It should be noted that the Rhines proof given by \citet{book:Salmon:1998}  is different from the proof given in the original papers \citep{article:Rhines:1975,article:Rhines:1979}. The difference is that in  \eqref{eq:rhinesass} $K$, which is time dependent, is replaced with a constant wavenumber $k_1$ representing the initial position of the peak. This modified proof was extended to the general case of $\ga$-turbulence by  \citet{article:Vallis:2002}. However, we feel that the original assumption \eqref{eq:rhinesass}  is more reasonable, on physical grounds, and  there is no benefit in  modifying \eqref{eq:rhinesass}.

Furthermore, it should be stressed that there is an important difference between the proof of \citet{article:Scott:2001} and the original Rhines proof. The main difference  is that Rhines assumes  that the \emph{unnormalized} variance of the energy spectrum is increasing  with time (Eq.\eqref{eq:rhinesass}) whereas \citet{article:Scott:2001} assumes that the \emph{normalized} variance $\gs_E^2$ is increasing.  The definition of $\gs_E^2$ is
\begin{equation}
\gs_E^2 \equiv \frac{\int_0^{+\infty}(k-K)^2 E(k) \; dk}{\int_0^{+\infty} E(k) \; dk}.
\end{equation}
Because the denominator of $\gs_E^2$ is decreasing with time, it is easy to see that the assumption $\dderivin{\gs_E^2}{t} > 0$ is mathematically weaker than the assumption \eqref{eq:rhinesass} used in the original formulation of the Rhines proof.  Consequently, since it is shown to be possible to arrive to the same conclusion under a weaker assumption, the statement proved  by \citet{article:Scott:2001} is better than the statement claimed by Rhines.  Thus, \citet{article:Scott:2001} implicitly also rehabilitates the original Rhines proof. However, \citet{article:Scott:2001} did not consider the case of hyperdiffusion and hypodiffusion in his paper.


For the more general case of hyperdiffusion and hypodiffusion, from the conservation laws, we find that  $E_0'$ and $E_2'$ read:
\begin{align}
E'_0 &= -2\nu E_{2p} - 2\nu_1 E_{-2h} \\
E'_2 &= -2\nu E_{2p+2} - 2\nu_1 E_{-2h+2},
\end{align}
and the time-derivative of $K^2$ is now bound by
\begin{align}
\frac{dK^2}{dt} &< \frac{E_2' -E_0' K^2}{E_0} = \frac{2\nu (K^2 E_{2p}  - E_{2p+2}) + 2\nu_1 (K^2 E_{-2h} - E_{-2h+2})}{E_0}
   \\
 &=\frac{2\nu}{E_0}\left(\frac{E_1^2 E_{2p}}{E_0^2} - E_{2p+2} \right) + \frac{2\nu_1}{E_0}\left(\frac{E_1^2 E_{-2h}}{E_0^2} - E_{-2h+2} \right) \\
 & = \frac{2\nu}{E_0}\frac{E_1^2 E_{2p}-E_0^2 E_{2p+2}}{E_0^2} + \frac{2\nu_1}{E_0}\frac{E_1^2 E_{-2h}-E_0^2 E_{-2h+2}}{E_0^2}.
\end{align}
The first term is again negative because 
\begin{align}
E_0^2 E_{2p+2} &= (E_0 E_{2p+2}) E_0 \geq (E_2 E_{2p}) E_0 = E_{2p} (E_0 E_2) \geq E_{2p}E_1^2,
\end{align}
for all $p>0$. Here, we employ the inequality $E_1^2 \leq E_0 E_2$, and the theorem that the function $\cE (\gk,\ga) \equiv E_{\gk+\ga}/E_{\gk}$ is an increasing function with respect to $\gk$ for $\ga>0$ (see appendix \ref{sec:holder}) which implies that $E_0 E_{2p+2} \geq E_2 E_{2p}$  for all real $p>0$. For $h=0$, the second term is negative too. To see this, note that the numerator of that term reads:
\begin{equation}
E_1^2 E_0 - E_0^2 E_2 = E_0 (E_1^2  - E_0 E_2) \leq 0.
\end{equation}
However, so far as we know, the sign of the second term is indeterminate when $0<h<1/2$ and can be shown to be positive when $h>1/2$. To show this, note that:
\begin{align}
\frac{E_0^2 E_{-2h+2}}{E_1^2 E_{-2h}} &\leq \frac{E_0 E_{-2h+1} E_{-2h+2}}{E_1 E_{-2h+2} E_{-2h}} =  \frac{E_0 E_{-2h+1}}{E_1 E_{-2h}} \leq \frac{E_{-2h} E_{-2h+1}}{E_{-2h+1} E_{-2h}} = 1,
\end{align}
Here we use $E_0/E_1 \leq E_{-2h+1}/E_{-2h+2}$, which is valid for $h>1/2$ and $E_0/E_1 \leq E_{-2h}/E_{-2h+1}$, which is valid for $h>0$. It follows that
\begin{equation}
E_1^2 E_{-2h}-E_0^2 E_{-2h+2} \geq 0, \text{ for } h >1/2.
\end{equation}
Thus, the validity of the Rhines proof continues for the case of hyperdiffusion ($p>1$), but cannot be extended to the case of hypodiffusion ($h>0$). 

It is interesting to note that when one begins with the weaker hypothesis of \citet{article:Scott:2001},  it can be shown that the contribution from the hypodiffusion term is \emph{always} positive for $h>0$. In the argument above we cannot show this unless $h>1/2$. To see this, we retrace the argument of \citet{article:Scott:2001} for the case of hyperdiffusion and hypodiffusion:
\begin{align}
\dderiv{K^2}{t} &= \frac{d}{dt}\fracp{E_2}{E_0} - \dderiv{\gs_E^2}{t} = \frac{E_2' E_0 - E_2 E_0'}{E_0^2}  - \dderiv{\gs_E^2}{t}\\
&= -\frac{2\nu}{E_0^2}(E_{2p+2} E_0 - E_2 E_{2p}) - \frac{2\nu_1}{E_0^2}(E_{-2h+2} E_0 - E_2 E_{-2h}) - \dderiv{\gs_E^2}{t}.
\end{align}
The first term is negative because $E_{2p+2} E_0 \geq E_2 E_{2p}$, for all $p>0$. As \citet{article:Scott:2001} noted, the second term vanishes for $h=0$, however it is \emph{positive} for $h > 0$. Thus, for  $h > 0$, the sign of $\dderivin{K^2}{t}$ remains indeterminate.

\section{Concluding remarks}

We have shown two inequalities \eqref{eq:ineqA} and \eqref{eq:ineqB}, which for the case of two-dimensional turbulence imply that the weighted-average of the energy flux is negative and the weighted-average of the enstrophy flux is positive. This implies that the energy tends to go upscale in the net and the enstrophy tends to go downscale in the net.  For the forced-dissipative case, the inequalities can be derived without any ad hoc assumptions.  For the decaying case, a sufficient condition for the energy inequality is to assume that there exists a very large wavenumber $k$ such that over the interval $(k,+\infty)$ the energy spectrum is  increasing or constant.  Likewise, for the enstrophy inequality it is sufficient  that we assume that there exists a very small wavenumber $k$ such that over the interval $(0,k)$ the energy spectrum is also  increasing or constant. From a physical point of view, these assumptions are slightly more plausible  than the assumption \eqref{eq:rhinesass} made by Rhines in his proof. It should be noted that unlike previous proofs in both the forced-dissipative and the decaying case, the inequalities have the same mathematical form. Our argument then is a unified proof that covers all cases, and specialized results can be deduced from our inequalities for special cases. We have also briefly discussed the implications of our results for one-layer and two-layer models of quasi-geostrophic turbulence.

Note that none of the results obtained in this paper forbids energy from being transferred downscale even when it is shown that the net flux should be directed upscale; they merely say that in those cases the energy going upscale in the upscale range should be larger than that going downscale in the downscale range.  In fact, for the case of finite domains with finite viscosity, \citet{article:Tung:2005,article:Tung:2005:1} showed that the downscale flux of energy on the short-wave side of injection must be nonzero.  Even in the case of 2-layer model where \citet{article:Orlando:2003} found in their numerical experiment that the downscale energy flux over the mesoscales contributes visibly to the observed energy spectrum, it is still true that there is a larger inverse energy cascade from the synoptic to the planetary scales.  The exception is the case of Surface QG turbulence, where most of the energy at the $z=0$ layer goes downscale, as shown here.  We suspect that this may be due to  the collapse of temperature gradients  on solid surfaces (a model of frontogenesis), and differs from the turbulence in the free atmosphere.  In the free troposphere, there is strong observational evidence (e.g. \citet{article:Shepherd:1983} and  \citet{article:Ditlevsen:1999}) that energy flux is negative (upscale) from synoptic to planetary scales, and the positive (downscale) flux over the mesoscales \citep{article:Lindborg:2001:1,article:Thornhill:2003} is small by comparison.

\begin{acknowledgements}
The research is supported by the National Science Foundation, under the grant DMS-03-27658. It is a pleasure to thank Sergey Danilov, Rob Scott, and an anonymous referee for their very helpful comments on our manuscript. 
 \end{acknowledgements}

\appendix
\section{H\" older inequalities}
\label{sec:holder}


Let $f(x)$ and $g(x)$ be two functions defined over a domain $x\in \cA$, such that
\begin{equation}
f (x) > 0 \text{ and } g(x) > 0, \;\forall x\in\cA,
\end{equation}
and let $a,b$ be real numbers such that $(1/a)+(1/b)=1$. Then the H\" older inequalities, in the integral form, read
\begin{equation}
\int_\cA f(x)g(x)\;dx \leq \left( \int_\cA [f(x)]^a \; dx\right)^{1/a} \left( \int_\cA [g(x)]^b \; dx\right)^{1/b}.
\end{equation}
For the case $a=b=2$ and $\cA = (0,+\infty)$ with $f(x) = k^{\ga}\sqrt{E(k)}$ and $g(x) = k^{\gb}\sqrt{E(k)}$, we have
\begin{align}
E_{\ga+\gb} &= \int_0^{+\infty} k^{\ga+\gb} E (k) \; dk \\
& \leq \left( \int_0^{+\infty} (k^{\ga}\sqrt{E(k)})^2 \; dk \right)^{1/2} \left( \int_0^{+\infty} (\sqrt{k^{\gb} E(k)})^2  \; dk \right)^{1/2} = \sqrt{E_{2\ga} E_{2\gb}}. 
\end{align}
For the cases $(\ga,\gb)=(0,1)$ and $(\ga,\gb)=(0,2)$ we obtain $E_1^2 \leq E_0 E_2$ and $E_2^2 \leq E_0 E_4$ by raising squares, noting that all the quantities involved are positive. For the case $\ga \mapsto \gk$ and $\gb \mapsto \gk+2\ga$ we get the inequality $E^2_{\gk+\ga} \leq E_\gk E_{\gk+2\ga}$ which can be rewritten as 
\begin{equation}
\frac{E_{\gk+\ga}}{E_{\gk}} \leq \frac{E_{\gk+2\ga}}{E_{\gk+\ga}}.
\label{eq:incrholderlemma}
\end{equation}
This inequality appears to indicate that the function $\cE (\gk,\ga) \equiv E_{\gk+\ga}/E_{\gk}$ is an increasing function with respect to $\gk$ for $\ga >0$ and decreasing for $\ga <0$. To prove this, we first note that from Eq.\eqref{eq:incrholderlemma} we have, for any integer $n >0$
\begin{align}
\cE (\gk +\ga/n, \ga) &= \frac{E_{\gk +\ga/n + \ga}}{E_{\gk +\ga/n}} = \prod_{j=1}^n \frac{E_{\gk +\ga/n + j\ga/n}}{E_{\gk +\ga/n + (j-1)\ga/n}} \\
&\geq \prod_{j=1}^n \frac{E_{\gk  + j\ga/n}}{E_{\gk + (j-1)\ga/n}} = \frac{E_{\gk+\ga}}{E_\gk} = \cE (\gk, \ga).
\end{align}
It follows that since $\cE (\gk,\ga)$ is a differentiable function with respect to $\gk$, that $\pderivin{\cE (\gk,\ga)}{\gk}\geq 0$ for $\ga>0$ and $\pderivin{\cE (\gk,\ga)}{\gk}\leq 0$ for $\ga <0$, and thus $\cE (\gk,\ga)$ is an increasing function with respect to $\gk$ for $\ga>0$ and a decreasing function for $\ga <0$. All the inequalities needed for our discussion of Rhines proof can be deduced from this result.

\bibliography{references,references-submit}
\bibliographystyle{jfm}

\end{document}